\documentclass[journal,review,submit,moreauthors,pdftex]{Definitions/mdpi} 
\firstpage{1} 
\pubvolume{1}
\issuenum{1}
\articlenumber{0}
\pubyear{2021}
\copyrightyear{2020}
\datereceived{} 
\dateaccepted{} 
\datepublished{} 
\hreflink{https://doi.org/} 

\renewcommand{\linenumbers}{}

\usepackage[T1]{fontenc}
\usepackage[utf8]{inputenc}

\usepackage{lastpage}
\usepackage{bm}
\usepackage{comment}
\usepackage{multicol}
\usepackage{multirow}
\usepackage{booktabs}
\usepackage{ulem}
\usepackage{times,amsmath,amssymb,longtable}
\usepackage[varg]{txfonts}
\usepackage{comment}
\usepackage{hyperref}
\usepackage{natbib}

\usepackage{grffile} 
\usepackage{graphicx}

\usepackage{color}

\newcommand\g {$\gamma$}
\newcommand\arcsec{^{\prime\prime}}

\Title{The Crab Pulsar and Nebula as seen in gamma-rays}
\TitleCitation{Crab in gamma-rays}

\Author{Elena Amato $^{1,2}$\orcidA{}, Barbara Olmi $^{1,3}$\orcidB{}}

\AuthorNames{Elena Amato, Barbara Olmi}

\AuthorCitation{Amato, E.; Olmi, B.}

\address{$^{1}$ \quad INAF - Osservatorio Astrofisico di Arcetri, Largo E. Fermi, 5, 50125, Firenze, Italy \\
$^{2}$ \quad Universit\`a degli Studi di Firenze, Via Sansone 1, 50019, Sesto Fiorentino (FI), Italy\\
$^{3}$ \quad INAF - Osservatorio Astronomico di Palermo, Piazza del Parlamento 1, 90134 Palermo, Italy}

\corres{Correspondence: elena.amato@inaf.it}

\abstract{Slightly more than 30 years ago, Whipple detection of the Crab Nebula was the start of Very High Energy gamma-ray astronomy. Since then, gamma-ray observations of this source have continued to provide new surprises and challenges to theories, with the detection of fast variability, pulsed emission up to unexpectedly high energy, and the very recent detection of photons with energy exceeding 1 PeV. In this article we review the impact of gamma-ray observations on our understanding of this extraordinary accelerator.}
\begin{document}
\section{Introduction}
\label{sec:intro}
The remains of the Supernova explosion AD 1054 are likely the best studied astrophysical system after the Sun \citep{hester:2008}. The remnant consists of two different bright non-thermal sources: the pulsar and the nebula. Both objects have played a key role in the development of high energy astrophysics. Thanks to their bright emission at all wavelengths, they have been observed by virtually all new astronomical instruments and have been at the origin of a wealth of important scientific discoveries. 

The Crab pulsar was one of the first detected pulsars and actually the one that provided smoking gun evidence for the identification of these radio sources as neutron stars. The Crab nebula had long been known to be the result of a SN explosion \citep{Lundmark:1921}; in 1934 \citet{Baade:1934} had suggested that supernova explosions might be signaling the transformation of an ordinary star into a neutron star, but the prospects for revealing these objects (small and presumably very dim) had been considered poor; in 1967 \citet{Pacini67} had suggested that a fast spinning, highly magnetized neutron star could be the energy source powering the activity of the Crab Nebula; in 1968 the first pulsar had been discovered and suggested to be a white dwarf or a neutron star \citep{HewishBell68}. The discovery of pulsations from one of the two stars at the center of the Crab nebula \citep{Staelin:1968} served as the last piece of the pulsar puzzle.

The contribution of the Crab pulsar and nebula to the progress of science did not end there, however. It is from this system that we have learned the basic physics behind the energy release by a young neutron star: the star spins down due to the electromagnetic torque and most of its rotational energy goes into the production of a relativistic magnetized wind. If this wind is effectively confined, as is the case for the Crab pulsar, the neutron star energy becomes detectable in the form of non-thermal emission by a surrounding nebula, the Pulsar Wind Nebula (PWN hereafter). This class of sources, of which the Crab nebula is the prototype, has typically a very broad non-thermal spectrum, often extending from low radio frequencies (tens of MHz) to Very High Energies gamma-rays ($E>100$ GeV photons; VHE hereafter). In fact, they account for the  majority of galactic sources emitting TeV gamma-rays, and also a number of unidentified gamma-ray sources are likely to be associated with unobserved pulsars \citep{HESS-GPS-2018}. 
Finally, very recent measurements by the LHAASO telescope \citep{cao2021peta} might indicate that PWNe are also the most numerous class of Extremely High Energies ($E>100$ TeV photons; EHE hereafter) gamma-rays emitters.

\end{paracol}
How exactly the star rotational energy is converted into the wind energy, and what the composition of the wind is, are questions with only partial answers. At the same time, the importance of these questions goes beyond pulsar physics, and, as we will discuss in this article, has implications for our understanding of particle acceleration in extreme conditions and up to the highest achievable energies, and on the origin of cosmic rays. Gamma-ray emission offers a privileged window to investigate these questions.  

On the other hand, gamma-ray observations of the Crab pulsar and nebula have kept surprising us with unpredicted discoveries, such as pulsations extending to unexpectedly high energies, extremely fast variability at GeV energies and detection of photons at PeV energies. In the following we discuss these discoveries and their implications for our understanding of pulsars, of the physics of relativistic plasmas and of particle acceleration up to the highest energies.  
The article is structured as follows. In \S~\ref{sec:PSR} we review our present understanding of
the properties of pulsar magnetospheres, with particular reference to the implications for pair production that come from the detection of VHE pulsed emission. In \S~\ref{sec:PWN} we review how the modeling of the nebular plasma has evolved, pushed by the improvement of observational capabilities at increasingly high energy. In particular,
we illustrate how 3D MHD modeling guided by high resolution X-ray data has affected our understanding of the wind properties and estimates of its parameters, and the kind of information that gamma-rays can provide. In \S~\ref{sec:TimVar} we 
discuss the problems at explaining particle acceleration in the Crab nebula and the insight that can be gathered from modeling the time variability of the source. 
The two major surprises that observations of the Crab nebula have offered us in recent years are presented in  \S~\ref{sec:flare} and \ref{sec:PeV}:
the gamma-ray flares and the detection of PeV emission. In \S~\ref{sec:Others} we discuss in what respects the Crab nebula is different from most of other objects in this source class, and how these differences might reflect in gamma-rays. Finally, we provide our summary and outlook in \S~\ref{sec:sumEprosp}.

\section{The Crab pulsar in gamma-rays: origin of the emission and pair multiplicity}
\label{sec:PSR}
As mentioned above, the Crab pulsar is a source whose existence had been predicted even before discovery \citep{Pacini67}, based on the need for an energy source to power the Crab nebula. Indeed, most of the pulsar spin-down energy, $\dot E\approx 5 \times 10^{38}$ erg s$^{-1}$, ends up in a magnetized wind expanding with relativistic bulk speed. At some distance from the star, the wind is slowed down to match the conditions of non-relativistic expansion of the conducting cage of supernova ejecta that confines it. This transition is thought to occur at a termination shock (TS hereafter), where the bulk energy of the outflow is dissipated and particles are accelerated, giving rise, thereafter, to the bright non-thermal nebula. We will worry about the bulk of the energy and address the nebular emission later in this article, while this section is devoted to the $\sim 1\%$ of $\dot E$ that goes into direct electromagnetic radiation, with a non-negligible fraction emitted in gamma-rays \citep{Harding:2013}.

The Crab pulsar is the source in this class with the broadest detected emission spectrum, extending from a few $\times 100$ MHz to TeV photon energies \citep{zanin:2017}. While the advent of {\it Fermi}-LAT has revealed that High Energy ($100-300$ MeV photons; HE hereafter) gamma-ray pulsations are not uncommon among pulsars \citep{Fermi2PC}, despite recent efforts \citep{VERITASPulseSearch}, no other pulsar has been firmly detected at VHE. The detection of the Crab pulsar in gamma-rays of progressively higher energy has had a tremendous impact on our ideas about pulsar magnetospheres and the mechanisms behind their emission in the different wavebands.

In spite of the fact that pulsars were first recognized as pulsating radio sources (to which they actually owe their name), and only later identified at shorter wavelengths, pulses of radio emission have always been the most challenging to account for in terms of theory, due to the coherent nature of their emission (see \citep{Melrose2017} for a review and \citep{Philippov20,Melrose21} for recent work on the subject). On the other hand, higher energy emission, from infrared frequencies upwards, is not coherent and has always appeared easier to understand as the result of classical emission processes, such as synchrotron, curvature and/or inverse Compton (IC) radiation, depending on the frequency and on the model. While near infrared through optical-UV - and often also non-thermal X-ray - emission is commonly accepted to be of synchrotron origin (see e.g. \citep{Romani96}), the process behind gamma-ray emission has long been debated \citep{Caraveo14}. 
Different emission mechanisms and different regions of origin are assumed by the different models, and in fact gamma-ray emission has long been thought to hold the key to understanding the hidden workings of the star magnetosphere \citep{Arons96}. 
Indeed, fundamental constraints have come from gamma-ray observations, especially in the VHE range.

The general picture of the pulsar immediate vicinities is thought to be as follows. A pulsar is a highly magnetized and fast spinning excellent conductor. While inside it charges organize themselves so as to screen the electric field, the unscreened field at the surface is strong enough to extract electrons and possibly even ions from the star, generating a co-rotating magnetosphere around the star \citep{GJ:1969}. 
The co-rotating magnetosphere can only extend up to a distance from the pulsar such that corotation does not imply superluminal motion: this defines the \textit{light cylinder} radius $R_{LC}=cP$, with $c$ the speed of light and $P$ the star rotation period. Magnetic field lines originating close enough to the pulsar magnetic axis (the so-called polar cap region) will not close within $R_{LC}$ and will form the open magnetosphere. Particles flowing along these lines meet regions of unscreened electric potential where they are accelerated and emit high energy radiation that subsequently leads to pair production.
It is through this process that each electron extracted from the star gives rise to $\kappa$ electrons, with $\kappa\gg1$ the so-called pulsar multiplicity. 
The open field lines are finally loaded with orders of magnitude more particles than originally extracted from the star surface: these particles flow away from the pulsar carrying with them most of the star rotational energy in the form of a magnetized relativistic wind, as we discuss further below. 
The exact multiplicity, {\it i.e.} the exact amount of pair production that should be expected from the magnetosphere of a given pulsar, is still a controversial subject (e.g. \citep{Timokhin19}). 
A way to estimate $\kappa$ from observations is by observing and modeling the PWNe, when possible. However, even in the case of the Crab nebula, the results obtained from this kind of observations are controversial, as we will discuss in more detail later in this article. Alternative constraints on the magnetospheric models and on the amount of pairs they produce can be derived from gamma-ray observations.

The big expectation in terms of the information that pulsed VHE emission might hold relates exactly to the topic of pair production. Particles extracted from the star quickly accelerate during the extraction process and emit high energy photons. In the intense magnetic field close to the star, photons with sufficiently large energies are absorbed and initiate a pair production cascade. 
The threshold energy for photons to escape rather than be absorbed, and give rise to a new generation of pairs, depends on the magnetic field strength, and therefore it will be different at different locations in the magnetosphere. This is why the detection of high energy gamma-rays was long awaited as a probe of the location of cascade development, and pair emission process. For the former, three main possible locations have been suggested since the early times of pulsar studies, the polar caps
\citep{Sturrock:1971,Ruderman&Sutherland:1975} the slot gaps \citep{Arons:1983,Muslimov&Harding:2004} and the outer gaps \citep{Cheng:1986}. In the first model, gamma-ray emission would come from the pulsar vicinity and should show a super-exponential cut-off at $\sim$ GeV energies, while in the latter two it would come from larger distances from the pulsar and be the result of curvature or Inverse Compton radiation, rather than synchrotron.

In addition to gap models, another scenario that satisfies this constraint is one in which particle acceleration and subsequent gamma-ray emission occurs in the equatorial current sheet of the pulsar wind, as a consequence of magnetic reconnection in the striped wind, taking place at distances from the pulsar comparable to $R_{LC}$ or larger \citep[e.g.][]{Kirk:2002}. If this process occurs close to $R_{LC}$, for young and energetic pulsars, such as Crab, it can come with associated pair creation: accelerated particles emit synchrotron gamma-ray photons which may create pairs through \g -\g\  interaction \citep{Lyubarskii:1996}.

The detection by {\it Fermi} of a large number of gamma-ray pulsars immediately seemed to disfavour polar caps as the main site of gamma-ray emission \citep{Watters&Romani:2011}: the simplest argument in this sense is the large number of detected pulsars, easier to reconcile with the wider beam of radiation predicted by models locating the emission further from the pulsar. More stringent constraints came from the detection of VHE pulsations from the Crab pulsar by MAGIC \citep{CrabMAGIC08,CrabMAGIC11} and VERITAS \citep{CrabVERITAS}: starting from 2008 the two telescopes detected pulsed emission from Crab at progressively higher energy, with the current record being 1.5 TeV \citep{MAGICcoll:2016}. 

These data enforce the view that gamma-ray emission comes from distances of order $R_{LC}$ or larger, with VHE gamma-rays most likely resulting from IC scattering of lower energy photons. 
At lower gamma-ray energies, the physical mechanism behind the emission is still debated between curvature \citep{PSRFundPlane}, synchrotron \citep{Cerutti:2016}, and synchro-curvature \citep{Torres19}. 
The spatial location of the emission, however, seems better established. Indeed, in the last 15 years, there has been enormous progress also in terms of modeling the pulsar magnetosphere and in the detailed comparison between models and data. Numerical studies of the pulsar magnetosphere have been evolving from the force-free and full MHD regime towards global PIC simulations including pair creation (see \citep{Cerutti&Beloborodov:2017} for a review, and references therein for further details). These latter studies are clearly the frontier in a complex multi-scale problem such as that of the pulsar magnetosphere. The general consensus is that, whenever the pair supply is sufficient to screen the electric field, the magnetosphere is globally well described by the force free solution (\citep{Spitkovsky:2006} and references therein), with the formation of a Y point near the light cylinder, where the equatorial current sheet connects with the two curved current sheets that form along the separatrix between open and closed field lines. In this case different prescriptions about the location of pair creation lead to similar results \citep{Philippov:2014, Philippov:2015,Brambilla:2018}. This is expected to be the case for young, fast spinning pulsars \citep{Chen:2014}, such as the Crab and most gamma-ray emitting pulsars. For these objects, current numerical simulations predict, in fact, that most of the high energy radiation results from synchrotron emission in the vicinity of the light cylinder \citep{Lyubarskii:1996,Kirk:2002,Cerutti:2016}. In the case of the Crab pulsar, this idea also gains support from the fact that detailed modeling of the light curve and optical polarization \citep{Cerutti_Mortier:2016} leads to determine values of the inclination between the pulsar magnetic and rotation axis and of the viewing angle that are in agreement with estimates based on completely different considerations related to the morphology of the nebula in X-rays \citep{Ng&Romani:2004}. 

The VHE emission from the Crab pulsar has never been computed within the refined global approach to magnetospheric dynamics and emission modeling discussed above. 
However, phenomenological modeling of phase-resolved spectra above 60 MeV \citep{Yeung:2020} strongly suggest that emission above 60 GeV comes from regions near or even beyond the light cylinder. In addition, even before the detection of pulsed TeV radiation, \citet{Mochol&Petri:2015} predicted multi-TeV gamma-rays as a distinctive signature of gamma-ray production via synchrotron-self-Compton at tens of $R_L$. 

\section{The Crab nebula: what we learn from gamma-rays}
\label{sec:PWN}
The Crab nebula has been known as a source of VHE gamma-rays since the late '80s \citep{Weekes:1989}, and was detected, for the first time, at MeV photon energies in the early '90s \citep{CrabEGRET}. The observed emission was readily interpreted as the result of IC scattering between the relativistic leptons populating the nebula and ambient photons, mainly contributed by the cosmic microwave background (CMB), thermal dust emission, and nebular synchrotron emission \citep{de-Jager:1992,Atoyan:1996}. 

\begin{figure}
\centering
	\includegraphics[width=.7\textwidth]{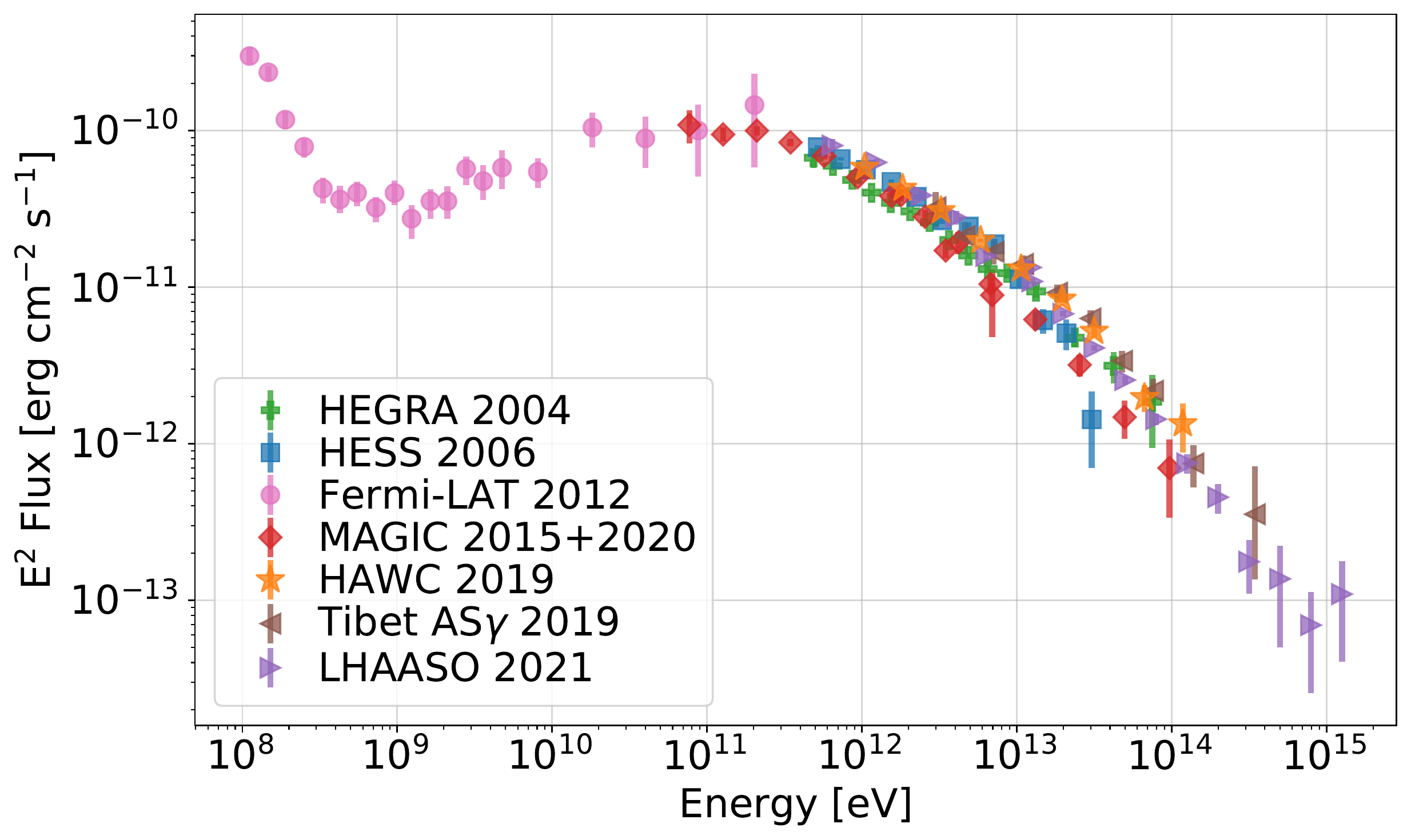}
    \caption{Focus on the gamma-ray spectrum of the Crab nebula. Data from different instruments are shown with diverse symbols/colors, namely: green rectangles for HEGRA data \citep{HEGRA:2004}, blue squares for HESS data \citep{Hess:2006}, pink circles for {\it Fermi}-LAT ones \citep{Buehler:2012}, red diamonds for MAGIC data \citep{MAGICcrab:2015, MAGICcrab:2020}, orange stars for HAWC \citep{HAWCcrab:2019}, brown triangles for Tibet AS-$\gamma$ \citep{tibetASg:2019} and violet ones for LHAASO data \citep{LHAASOcrab:2021, cao2021peta}. Figure courtesy of Michele Fiori.}
    \label{fig:CRABspectrumIC}     
\end{figure}
In the last 15 years, the advent of the current generation of HE ({\it Fermi}-LAT and AGILE) and VHE (MAGIC, VERITAS, H.E.S.S., HAWC, Tibet As-\g, LHAASO) gamma-ray telescopes has allowed to gain much deeper insight in the properties of the Crab nebula at these highest energies, and has also brought two big surprises: variability in the MeV range \citep{AGILEFlare,FermiFlare} and detection up to unexpectedly high energies \citep{CrabLHAASO}. In Fig.~\ref{fig:CRABspectrumIC} we show the most recent measurements of Crab nebula gamma-ray spectrum, including LHAASO data points, showing emission beyond 1 PeV, about the highest energy we think achievable by galactic accelerators, based on measurements of the cosmic ray spectrum at the Earth (see e.g. \cite{AmatoCasanova} for a recent review). Before discussing the most impressive surprises that came from gamma-rays and how they impact our understanding of the Crab nebula, we briefly review the physical picture of the nebular dynamics and emission properties that has been built through time, thanks to constant improvement of the quality of observations, of theories and numerical modeling.

\subsection{Modeling the nebular plasma}
\label{sec:MHD}
The Crab nebula is the PWN for which most models were developed and over which most of our understanding of the entire class is based. As we mentioned in \S \ref{sec:intro} most of the rotational energy lost by the pulsar goes into accelerating a relativistic outflow, mostly made of pairs (though the presence of ions is not excluded, as we will discuss later) and a toroidal magnetic field.
The outflow starts out cold (low emissivity, as highlighted by the presence of an under-luminous region surrounding the pulsar \citep{hester:2008}) and highly relativistic, until it reaches the termination shock (TS). Since the outflow is electromagnetically driven, it must start out as highly magnetized at $R_{LC}$: the ratio between Poynting flux and particle kinetic energy, $\sigma$, is thought to be $\sigma(R_{LC})\approx 10^4$ \citep{KirkLyub09,Arons12}. In contrast, the magnetization must be much lower at the TS, in order for the flow to be effectively slowed down. Initial estimates of $\sigma$ at the TS, based on steady state 1D magnetohydrodynamics (MHD) modeling would give $\sigma(R_{TS})\approx 10^{-3}$, equal to the ratio between the nebular expansion velocity and the speed of light. This estimate has later been revised towards larger values of $\sigma$ in light of 3D MHD numerical modeling, as we discuss below, but the general consensus is still that $\sigma(R_{TS})$ cannot be much larger than unity. How the conversion of the flow energy from magnetic to kinetic occurs, between $R_{LC}$ and $R_{TS}$, is still a matter of debate -- the so-called $\sigma$-problem -- and some of the suggested mechanisms could show radiative signatures in the gamma-ray band ({\it e.g.} \cite{Kirk:2002}), while keeping dark in other wavebands. In fact, at least at low latitudes around the pulsar rotational equator, a plausible mechanism for energy conversion in the wind is offered by the existence of a magnetically striped region \citep{Coroniti90}. In an angular sector, whose extent depends on the inclination between the pulsar spin and magnetic axes, $\theta_i$, a current sheet develops between toroidal field lines of alternating polarity \citep{Spitkovsky:2006}: this is an ideal place for magnetic reconnection to occur and transfer energy from the field to the plasma \citep{Coroniti90}. Where along the flow and whether efficiently enough this energy conversion occurs is an open question, the answer to which depends on the pair-loading of the flow \citep{LyubKirk01}, namely on the pulsar multiplicity $\kappa$, again a parameter to be preferentially investigated in gamma-rays.
This latter statement is true in two respects: constraints on pair-production in the magnetosphere can be gained from pulsed gamma-ray emission, as discussed in \S~\ref{sec:PSR}, but a more direct estimate of the number of pairs injected in the nebula can be obtained from detailed modeling of the nebular emission spectrum and morphology. This is what we discuss in the following.

\begin{figure*}
\centering
	\includegraphics[width=.4\textwidth]{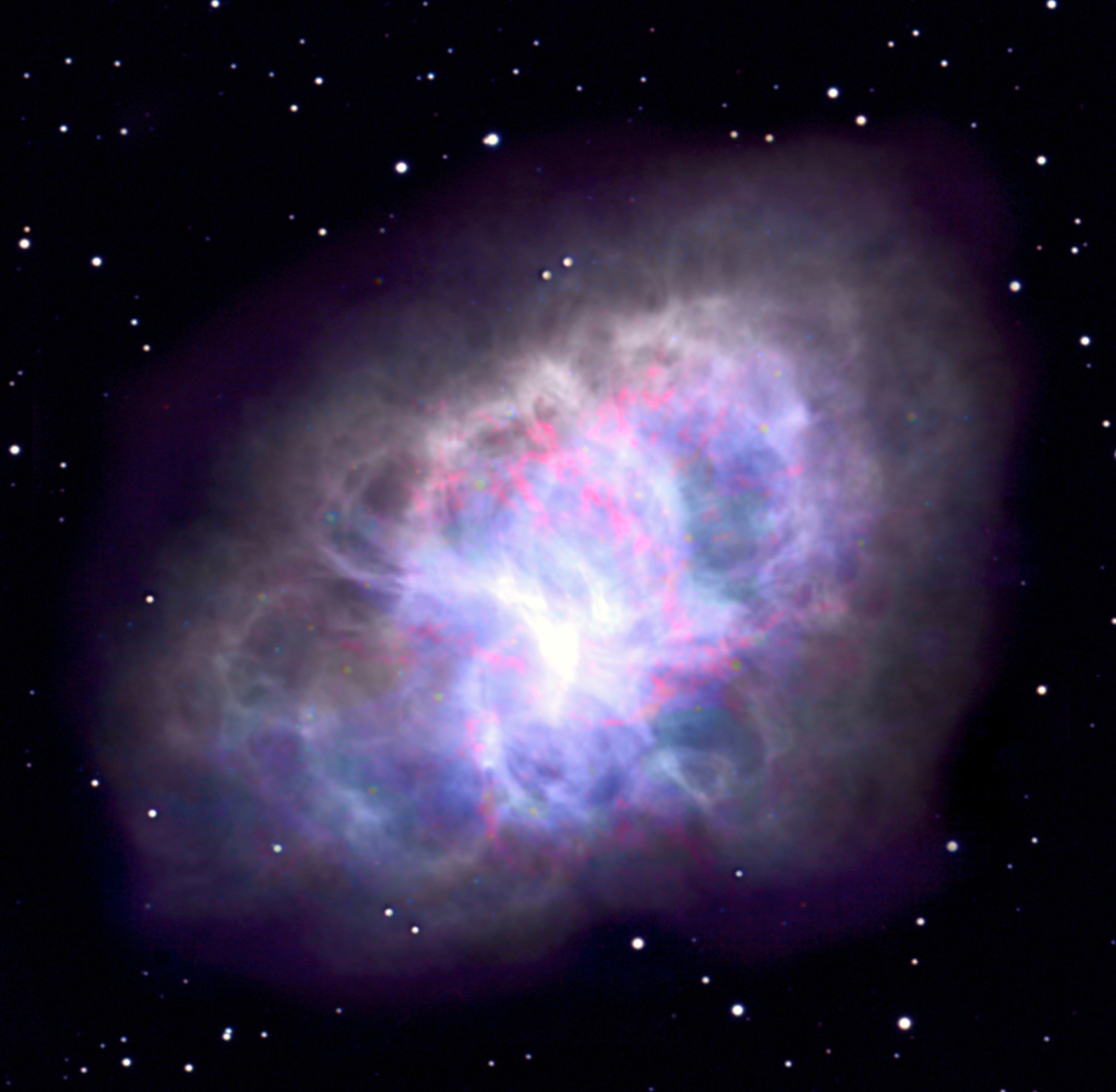}\,
	\includegraphics[width=.392\textwidth]{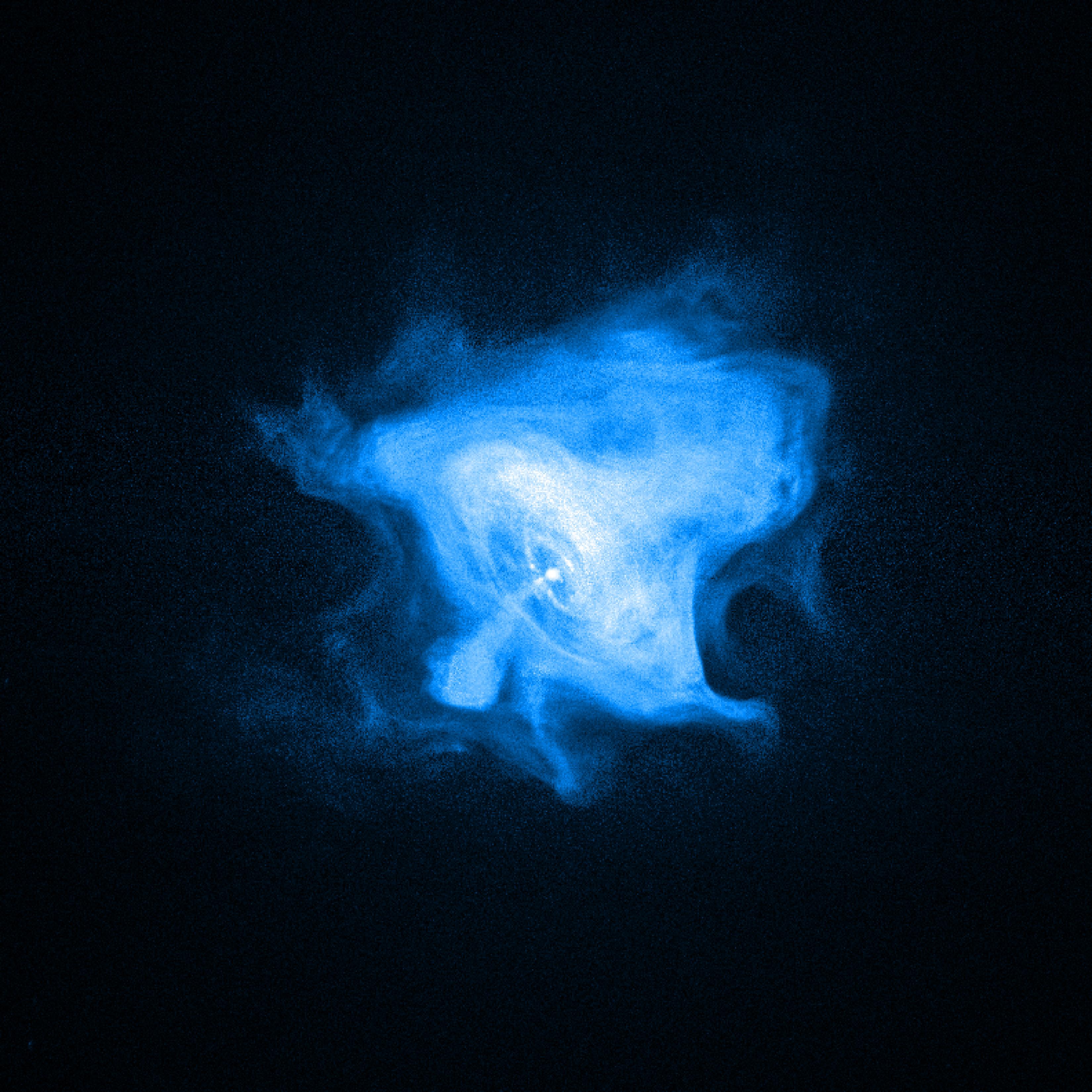}
    \caption{\textit{Left panel:} The Crab nebula as seen in radio with the National Radio Astronomy Observatory (credits: M. Bietenholz, T. Burchell NRAO/AUI/NSF; B. Schoening/NOAO/AURA/NSF).
    \textit{Right panel:} The Crab nebula in X-rays, as seen by  {\it Chandra} (credits: {\it Chandra} X-ray Observatory NASA/CXC/SAO/F.Seward et al.).}
    \label{fig:morphradX}     
\end{figure*}

The morphology of the synchrotron nebula is known in great detail, at photon energies from radio to X-rays (see Fig.~\ref{fig:morphradX}) and hence represents both a driver and a very challenging test for theoretical and numerical models. The size of the nebula is observed to vary noticeably with the energy of the emitting electrons, and consequently with the observation waveband. The higher the energy of the electrons is, the shorter the distance they travel before losing most of their energy due to synchrotron radiation. 

The most advanced available modeling of the Crab nebula so far is based on the assumption that beyond the TS, MHD provides a good description of the flow dynamics. 1D MHD models, both stationary and self-similar, were proposed since the '70s \citep{Rees_Gunn:1974, Kennel_Coroniti:1984a, Emmering_Chevalier:1987}, as well as stationary 2D solutions \citep{BegelmanLi92}. These models could generally account for the size shrinkage of the nebula with increasing frequency as a result of advection and synchrotron losses, for an average magnetic field in the nebula close to the equipartition value and for the synchrotron luminosity of the nebula, assuming a wind magnetization $\sigma\approx 3 \times 10^{-3}$, a wind Lorentz factor $\Gamma_w\approx 3 \times 10^6$ and an injection rate of particles in the nebula $\dot N\approx 10^{38}\ {\rm s}^{-1}$. Particles responsible for radio emission could not be accounted for with these values of the parameters.

The discovery by {\it Chandra} of a jet-torus morphology of the inner nebula \citep{Weisskopf:2000} prompted efforts to model the system with 2D axisymmetric MHD simulations, assuming a latitude dependence of the pulsar outflow \citep{Komissarov:2004,del-Zanna:2004}. The latter was taken in agreement with the split-monopole solution proposed by \cite{Michel:1973}, and later proved to provide a very good description of the force-free pulsar magnetosphere \citep{Spitkovsky:2006}: the pulsar wind flows along streamlines that become asymptotically radial beyond $R_{LC}$ and has an embedded magnetic field that is predominantly toroidal, with alternating polarity in a region $2 \theta_i$ around the equator. In this angular sector, magnetic dissipation is usually assumed to occur before the TS.

The energy flux in the wind has a latitude dependent distribution, with most of the energy concentrated in the pulsar equatorial plane. As a consequence, the pulsar wind TS does not have a spherical surface, but rather a highly oblate shape, being much closer to the pulsar along the rotational axis than at the equator. The obliquity of the shock front plays a key role to explain the X-ray observations of polar jets. These appear to originate so close to the pulsar position that, if the shock were spherical, they would have to be collimated directly in the highly relativistic plasma upstream of the shock, where known mechanisms are not efficient \citep{Lyubarsky:2002}. 2D MHD simulations proved that collimation happens, in fact, in the downstream plasma, as soon as magnetic hoop stresses are sufficiently strong, namely as soon as the magnetic field in the nebula can reach equipartition. This reflects in a lower limit on the wind magnetization for the jets formation: $\sigma\gtrsim10^{-2}$ \citep{Komissarov:2004,del-Zanna:2004,del-Zanna:2006}, about one order of magnitude larger than the value provided by 1D models. 

\begin{figure*}
\centering
	\includegraphics[width=.8\textwidth]{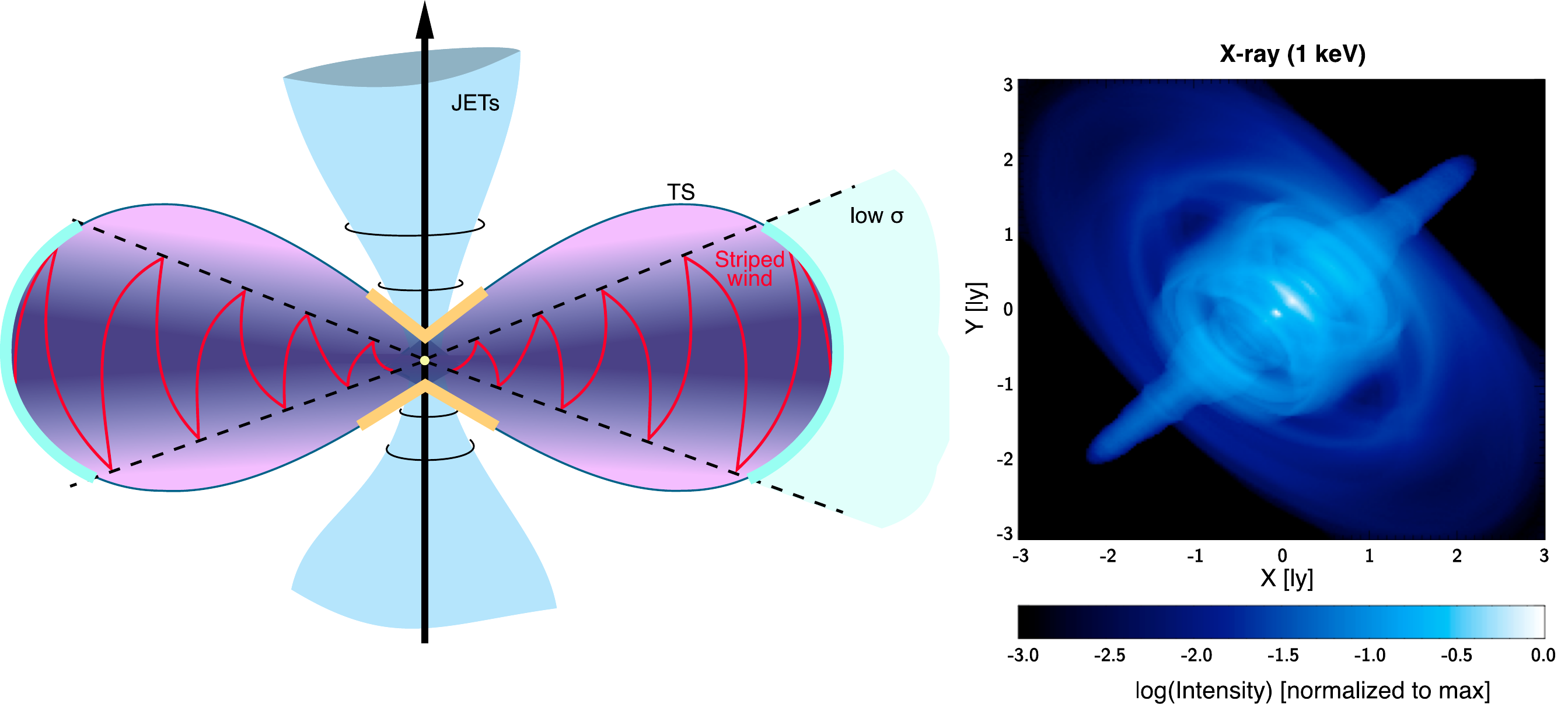} 
    \caption{\textit{Left panel:} Cartoon of the inner nebula geometry (the oblate TS, jets formation, striped wind) with the identification of the accelerating regions for particles responsible for the {\it wisps} emission at the different wavelengths. \textit{Right panel:} Surface brightness map at X-ray energies (1 keV), with intensity  normalized to the maximum value and expressed in logarithmic scale.\\ 
    Reprinted with permission from Del Zanna et al. (2006) $\copyright$ 2006 ESO.}
    \label{fig:cartoon}     
\end{figure*}

A schematic representation of the flow geometry can be seen in the left panel of Fig.~\ref{fig:cartoon}.In the right panel of the same figure, we show a simulated X-ray image of the Crab nebula.

2D axisymmetric models have proven very successful at accounting for the morphological properties of the Crab nebula emission. They very well reproduce most of the observed brightness features in the inner nebula in very fine detail, including the X-ray rings and the {\it knot} \citep{Weisskopf:2000,Lyutikov:2016}. On a larger scale, they account reasonably well for the shape of the nebula (elongated along the pulsar rotation axis \citep{Olmi:2014}) and for size shrinkage with increasing frequency, from radio to X-rays \citep{del-Zanna:2006}. 

\begin{figure*}
\centering
	\includegraphics[width=.8\textwidth]{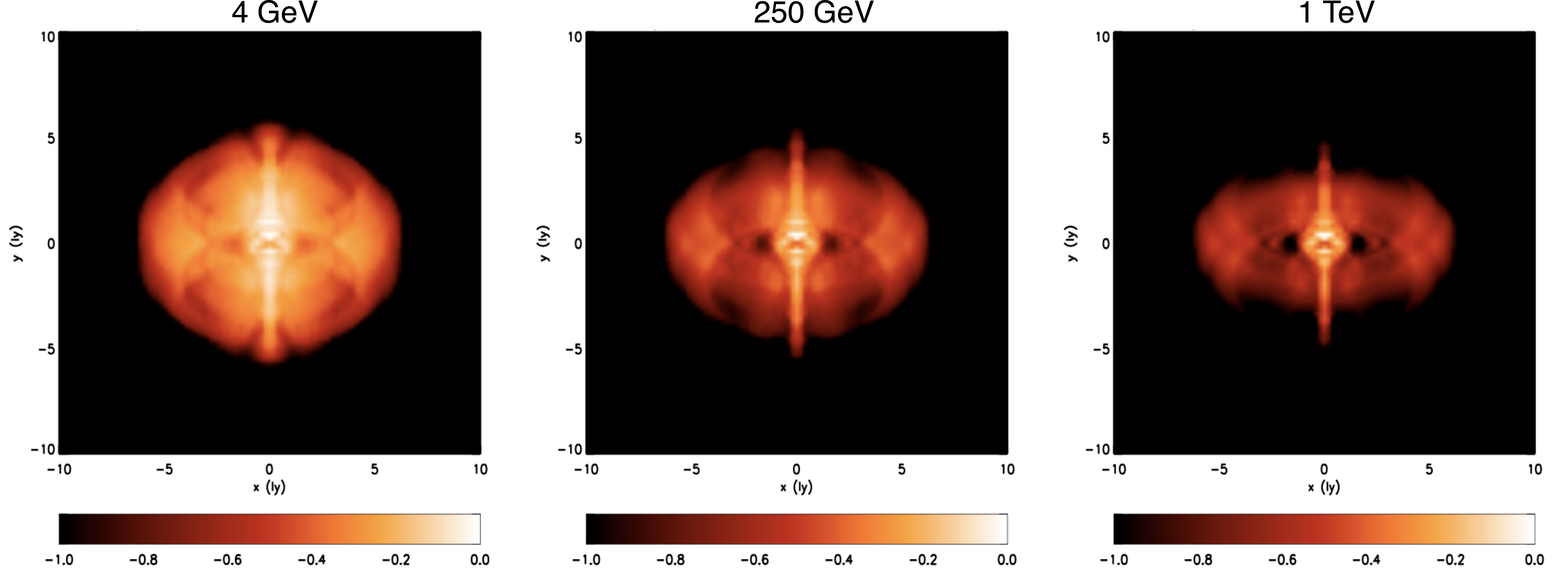} 
    \caption{IC surface brightness maps at various energies in the gamma-ray range. Each map is normalized to its maximum and plotted in logarithmic scale.\\
    Reprinted from Volpi et al. (2008) $\copyright$ 2008 ESO.}
    \label{fig:Volpi}     
\end{figure*}
As far as gamma-rays are concerned, no detailed morphological information is available, due to the very limited angular resolution of gamma-ray telescopes. For a long time the only available information simply constrained the gamma-ray nebula to lie within the radio synchrotron one \citep{Aharonian:2000,Albert:2008,Hess:2006}. The first direct measurement of the Crab nebula extension in gamma-rays became available last year, thanks to the H.~E.~S.~S. telescope \citep{HessNat:2020}. With the analysis of 22 hours of observations collected during 6 years of operation, the PWN radial extension was finally determined: it turns out to be $\sim 52\arcsec$ in the 700 GeV-5 TeV energy range, and hence smaller than in the UV (where the extension is $\sim 2.5^{\prime}$), and very similar to the X-ray size ($\sim 50\arcsec$), which is perfectly consistent with a picture in which TeV gamma-rays are produced by synchrotron X-ray emitting particles. This is also in very good agreement with the results of the only available effort at computing simulated gamma-ray emission maps of the Crab nebula \citep{Volpi:2008}. In this work the IC emission was computed on top of a 2D MHD numerical model and maps were produced for different photon energies, showing a shrinkage with increasing energy similar to that observed between radio and X-rays. This can be seen in Fig.~\ref{fig:Volpi}, which also clearly shows how the jet-torus structure should become visible again at TeV energies. Probing the nebular morphology at this level of detail in VHE gamma-rays is however beyond the reach of current and planned instruments \citep{Mestre:2020}.   
\begin{figure*}
\centering
	\includegraphics[width=.95\textwidth]{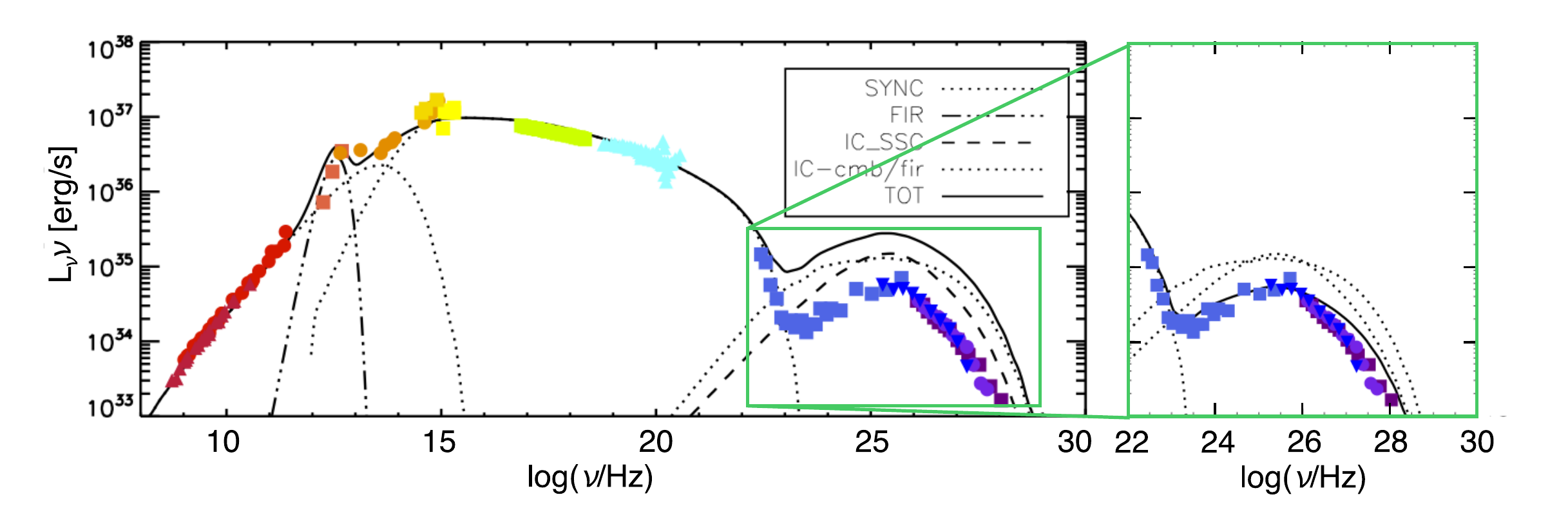} 
    \caption{Total integrated spectrum of the Crab nebula computed on top of the 2D MHD numerical model by \citep{Olmi:2014}. The zoom-in on the gamma-ray spectrum highlights the fact that the IC emission can be correctly reproduced if the magnetic field strength is artificially rescaled so as to ensure an average value of $\sim$ 200 $\mu$ G (this is how the spectrum in the inset is obtained).}
    \label{fig:crab_spectrum}     
\end{figure*}

One thing that gamma-rays can readily probe, however, is the goodness of 2D MHD models at correctly describing the energy content of the nebula: in fact, the main limitations inherent to the assumption of axisymmetry become apparent as soon as one compares the IC spectrum computed from simulations with the available data. As shown in Fig.~\ref{fig:crab_spectrum}, the 2D MHD simulations largely overpredict the IC flux. Indeed, the limits of axisymmetric models are evident when trying to describe the large scale properties of the PWN, primarily the global magnetic field structure. The imposed symmetry reflects in an artificial pileup of magnetic loops along the polar axis and an enhanced compression of the magnetic field in the inner nebula. In order to reproduce the nebular morphology one is then forced to adopt an artificially low magnetization of the flow ($\sigma\leq 0.1$), and as a result the overall magnetic energy in the nebula is underestimated. In order to reproduce the synchrotron spectrum, one is then forced to inject in the nebula a number of particles larger than in reality, which is readily revealed by the IC flux. The particle energy losses are also underestimated, and this forces one to assume an injection spectrum for high energy particles that is steeper than what is deduced from X-ray spectral index maps of the inner nebula \citep{del-Zanna:2006}.

The solution to many of these problems appeared with results from the first 3D MHD simulations \citep{Porth:2014}.
%
With the third spatial dimension available, kink-type plasma instabilities produce considerable mixing of the magnetic field in the entire nebula, with an ensuing high level of magnetic dissipation. This definitely allows for the increase of the initial magnetization in the pulsar wind to values of order of unity \citep{Porth:2014, Olmi:2016, Porth17, Olmi&Bucciantini:2019_1}. 
The main limitation of 3D models is that they require a huge amount of numerical resources and time to be performed. For this reason, in \cite{Porth:2014}, only a very initial phase of evolution of the Crab nebula was investigated, for a total of $\sim 70$ years, so that the self-similar expansion phase was not yet reached. A longer simulation, fully reaching the self-similar expansion phase, were presented in \cite{Olmi:2016}. Synchrotron emission maps computed on top of these simulations show that, for parameters appropriate to reproduce the X-ray morphology, the surface brightness distribution at radio and optical frequencies becomes much more uniform in 3D, reflecting the structure of the magnetic field, which appears to be rather different from what originally found based on 2D models \cite{Olmi:2015}, with differences increasing with distance from the shock and from the equatorial plane. 

\begin{figure}
\centering
	\includegraphics[width=.5\textwidth]{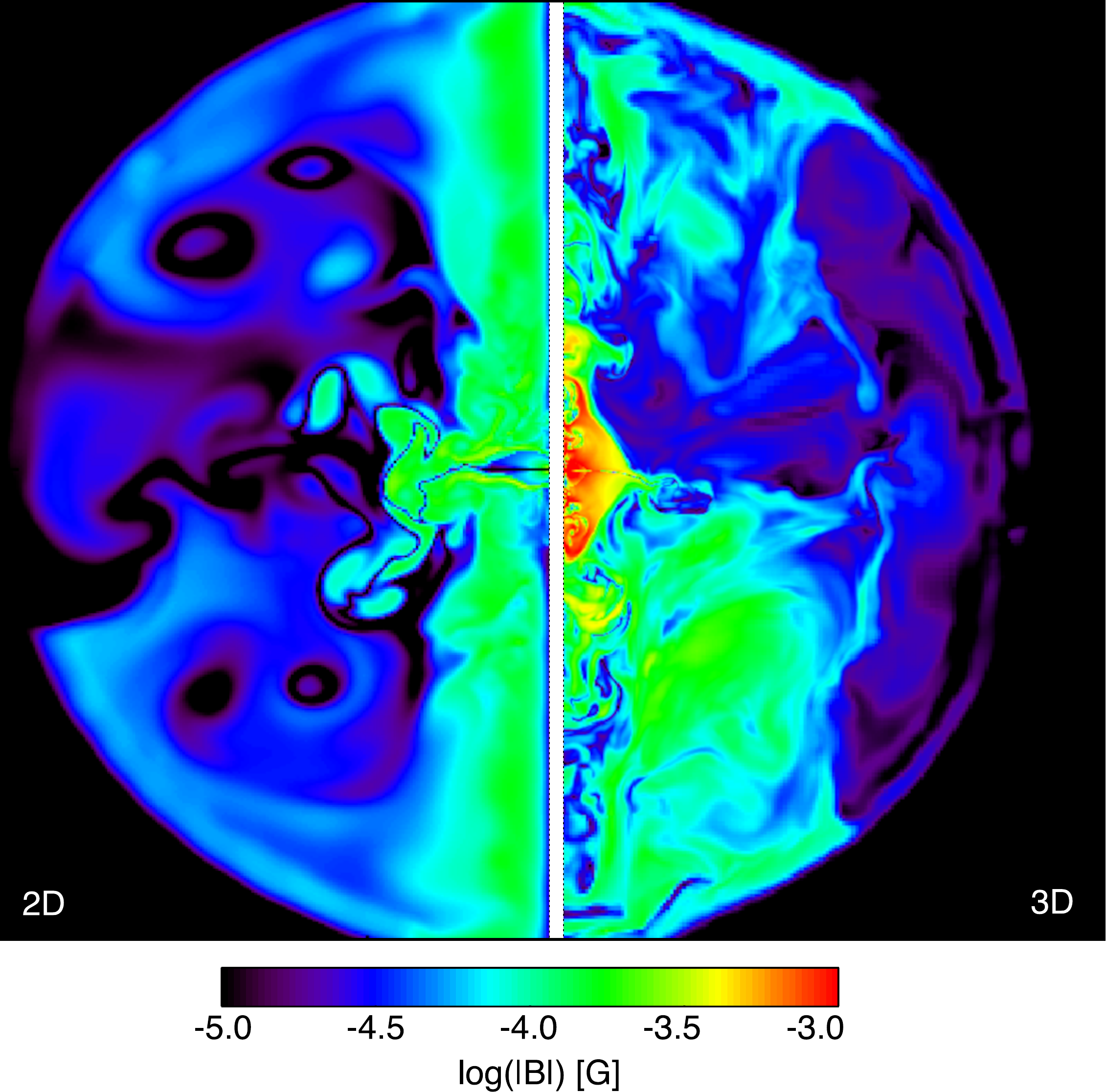} 
    \caption{Comparison of the magnetic field intensity (in logarithmic scale and units of G) between a 2D MHD model and a 3D one, that both reproduce the X-ray morphology (from original simulations presented in \cite{Olmi:2015,Olmi:2016}).}
    \label{fig:compareB}     
\end{figure}

In Fig.~\ref{fig:compareB} we show color maps of the magnetic field strengths in 2D (left) and 3D (right) corresponding to $\sigma=0.025$ and $\sigma=1$ respectively. The first thing to notice is that in 3D the pile-up of field lines around the polar axis is much reduced and their filling factor in the nebula much more uniform. This is due to the fact that, even injecting a toroidal magnetic field at the shock surface, the mixing is so efficient that a poloidal component immediately develops, becoming comparable in magnitude to the toroidal one within a distance from the pulsar of order 2-3 times the TS radius. On the other hand, the magnetic field remains almost toroidal in the inner nebula, making predictions from 2D axisymmetric models still valid limited to this region.

The second noticeable thing is that, in spite of the much higher magnetization adopted for the 3D simulation (a factor 40 larger $\sigma$), the average magnetic field in the nebula is only about a factor 2 higher than in 2D. This is a result of efficient magnetic dissipation: \cite{Olmi:2016} found that magnetic dissipation is so high that even an initial magnetization of order unity is not enough to lead to an average magnetic field of the expected strength order $\sim$ 150-200 $\mu$G, so that the actual wind magnetization might have to be even larger than unity, revising by more than 3 orders of magnitude the initial estimate based on 1D steady state modeling and strongly mitigating the $\sigma-$problem.

Before concluding this section, we think it is important to remark that in current 3D simulations magnetic dissipation has a purely numerical origin, while the actual physical process at work in the Crab nebula plasma remains unconstrained. In reality, how much of the injected toroidal field is left at any point in the nebula can be constrained by comparison of polarization maps with observations (see e.g. \citep{NicPolar2005,Porth:2014}). Important new insights in this respect will soon be provided by the availability of X-ray polarimetric observations \citep{XIPE13}.

\subsection{Time-variability and particle acceleration}
\label{sec:TimVar}
The era of multi-D MHD simulations also opened up the possibility of using spatially resolved time-variability as an additional, powerful diagnostic for the physical properties of the plasma in the nebula, and, most notably, also for the processes responsible for particle acceleration within it.
Brightness variations of the nebular structures had been known to occur, at optical frequencies, for a long time: the so-called {\it wisps} were first identified by \citep{Scargle69}. These features, strongly resembling  outward propagating plasma waves, appear at distances from the pulsar comparable with the TS radius in the equatorial plane and then progressively fade while moving outward, with time-scales from weeks to months \citep{Hester:2002}. Similar features were later observed both in the X-rays \citep{Weisskopf:2000} and in the radio band \citep{Bietenholz:2001,Bietenholz:2004}. In spite of these morphology variations, however, the integrated emission was found to vary only by a few percent per year \citep{Wilson-Hodge2011}.

The {\it wisps} appearance and time evolution however is not the same at all wavelengths \cite{schweizer13}, and varies in a way that, within the MHD framework, can only be interpreted as due to differences in the particle spectrum at different locations along the shock front, or, in other words, to particles in different energy ranges being accelerated in different places \cite{Olmi:2015}. On the other hand, the plasma conditions along the TS front are expected to be highly non-uniform, especially in terms of magnetization of the flow (see Fig.~\ref{fig:cartoon}), and this is an important parameter to determine the kind of acceleration process that can be locally at work, as we discuss below.

\begin{figure*}
\centering
	\includegraphics[width=.8\textwidth]{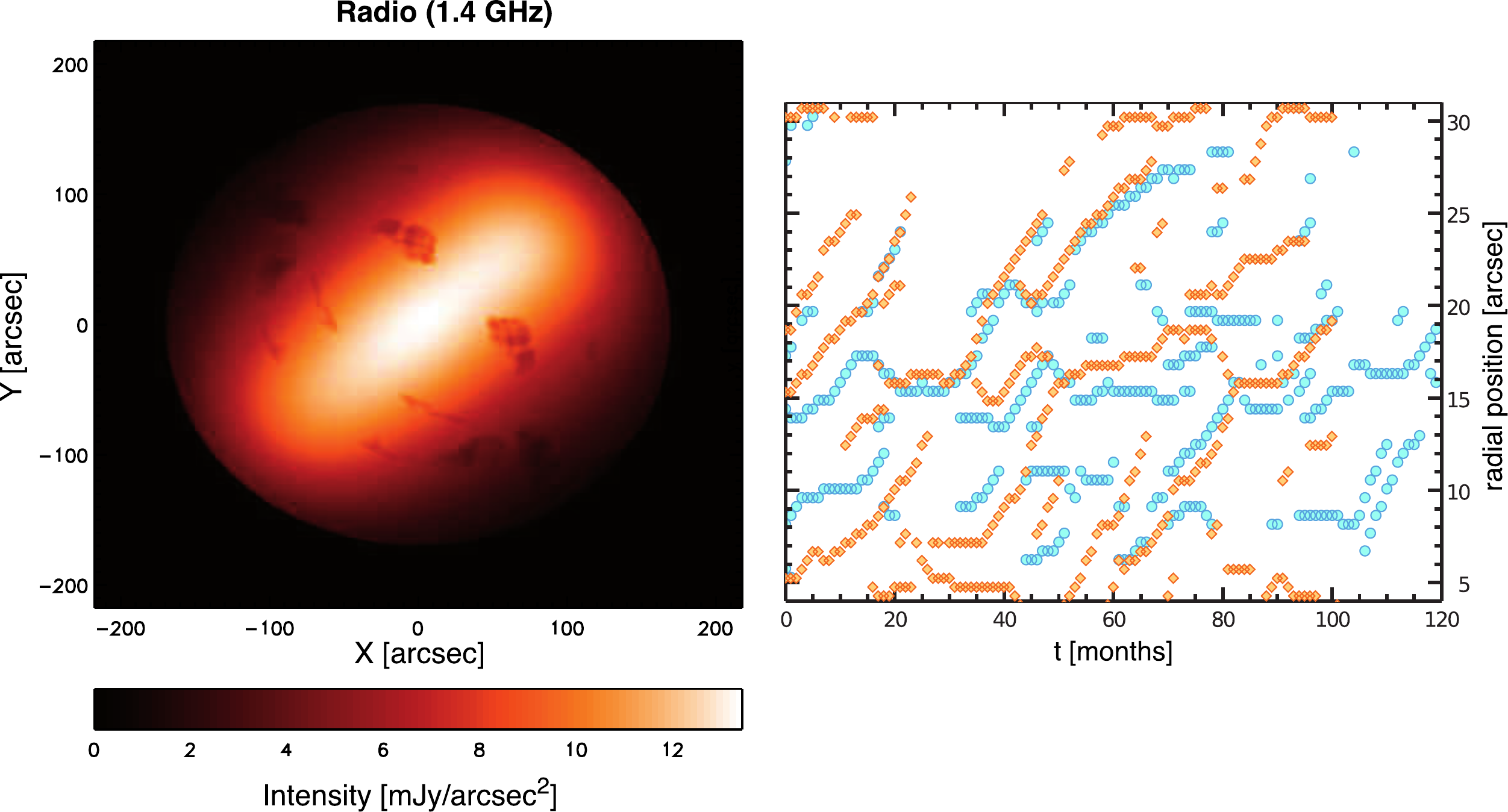} 
    \caption{{\textit{Left panel:} Surface brightness map at radio frequencies (1.4 GHz). Small scales have been subtracted and the map convolved with the VLA PSF. The intensity is given in linear scale and in mJy/arcsec$^2$ units. The emitting particles are assumed as uniformily distributed in the nebula.
    \textit{Right panel:} Non-coincidence of the X-ray (aquamarine circles) and radio at 5 GHz (orange diamonds) wisps, produced by particles accelerated in the regions highlighted with the same colors in the left panel of Fig.~\ref{fig:cartoon}. More discussion on this can be found in \citep{Olmi:2015}.\\
    The map in the left panel is reprinted with permission from Olmi et al. (2014) $\copyright$ 2014 Olmi et al.}}
    \label{fig:compareSYNC}     
\end{figure*}

In fact how particle acceleration occurs in the Crab nebula in different energy ranges is not understood (see e.g. \citep{Amato:2014,Amato19} for a review). The nebular synchrotron spectrum is consistent with a broken power-law, with a particle spectral index $\gamma_R=1.6$ for radio emitting particles and  $\gamma_X=2.2$ for X-ray emitting ones (see e.g. \citep{Amato2000}). At the highest energies particles must be accelerated at the TS, otherwise the decrease in size of the nebula with increasing frequencies could not be explained. On the other hand, radio emitting particles could be in principle accelerated anywhere in the nebula. 
The evidence of the co-existence of two different particle populations has been suggested by \citet{Bandiera:2002} after a comparison of radio, millimetric and X-ray maps of the Crab nebula.
The observation of {\it wisps} at radio frequencies seemed to exclude this possibility \citep{Bietenholz:2001}, but at a closer look this phenomenon can well be accounted for within the MHD framework as simply due to the structure of the magnetic field and of the MHD flow: \citep{Olmi:2014} showed that radio emission maps and time-variability can be reproduced even assuming that radio emitting particles are uniformly distributed in the nebula, as would be the case for diffuse acceleration in the body of the nebula, associated with stochastic magnetic reconnection or Fermi-II process due to MHD turbulence.
The frequency dependent behaviour of the {\it wisps} can only be accounted for, within MHD transport, if X-ray emitting particles are accelerated in the equatorial sector of the TS, while lower energy particles are predominantly accelerated elsewhere, either in the body of the nebula or at high latitudes at the TS \citep{Olmi:2015}. In Fig.~\ref{fig:compareSYNC} we show, on the left, the radio emission map obtained by \citep{Olmi:2014} assuming a uniform distribution of radio emitting particles in the nebula. The right panel of the same figure shows, instead, the time-evolution of the surface brightness peak at radio (orange) and X-ray (blue) frequencies when particles are injected in the sectors of the TS shown in Fig.~\ref{fig:cartoon}, highlighted with the corresponding colors.

With an estimated Lorentz factor of the wind in the range $10^4-10^7$, the shock in the Crab nebula is among the most relativistic in nature. The mechanism usually invoked for particle acceleration in astrophysical sources, diffusive shock acceleration or first order Fermi process (Fermi-I), can only work at such a shock if the magnetization of the wind is low enough, $\sigma\lesssim 10^{-3}$ \citep{Sironi:2009} (see \citep{SironiLemoine15} for a review). This condition can only be realized in a small equatorial sector of the wind, assuming efficient magnetic reconnection in the striped wind upstream of the shock, or in the vicinity of the polar axis, when the magnetic field naturally decreases and O-point type reconnection is also possible. The results found by \citep{Olmi:2015} concerning the preferred location of X-ray emitting particle acceleration are consistent with acceleration occurring mainly in the equatorial region and $\gamma_X$ is consistent with the outcome of Fermi-I acceleration. A question that remains open, and waits to be addressed in the framework of 3D MHD simulations, is whether a sufficiently large fraction of the flow satisfies the condition of low $\sigma$ required by the Fermi-I process.

Other possible acceleration mechanisms that have been suggested are associated with driven magnetic reconnection occurring at the TS \citep{Sironi:2011} or resonant absorption of ion cyclotron waves \citep{HoshinoArons,AmatoArons}. The former requires very large wind magnetization ($\sigma\gtrsim 30$ at the TS) and pair multiplicity ($\kappa\gtrsim 10^8$), while the latter requires the presence of ions in the pulsar wind. Both questions are again to be addressed by gamma-ray observations (see e.g. \citep{Amato19}). 

As far as requirements on $\kappa$ are concerned, from the point of view of pulsar theory, a value so large as $\kappa\approx 10^8$ seems very difficult to account for, in spite of the recent and ongoing evolution of pulsar magnetospheric models (section \S~\ref{sec:PSR}). In addition, with $\kappa\approx 10^8$ the wind would reconnect before the TS \citep{LyubKirk01} (with possible signatures in gamma-rays \citep{Kirk:2002}) and the magnetization could not be as high as required. Finally, even ignoring all the theoretical difficulties, and simply counting the number of particles that have accumulated in the nebula during its history, through combined modeling of the synchrotron and IC spectrum, that value of $\kappa$ is too large by $\approx$ 2-3 orders of magnitude \citep{Bucciantini11}. Of course the lack of evidence and/or motivation for large $\kappa$ does not exclude the possibility for magnetic reconnection to be responsible for acceleration in a limited energy range, as we further discuss in \S~\ref{sec:flare}.

Concerning acceleration via ion-cyclotron absorption, this mechanism requires a sizeable fraction of the wind energy to be carried by ions \citep{AmatoArons}, and hence that the pulsar multiplicity be not too large $\kappa\lesssim 10^4$ \citep{Amato19}. The implied population of ions would be made of particles with  Lorentz factor equal to that of the wind, $10^4<\Gamma_w<10^7$, and the only direct probe of their presence can come from gamma-ray or neutrino emission \citep{Amato:2003}. Recent LHAASO observation of the Crab nebula might hold important clues in this respect \citep{LHAASOcrab:2021}. This aspect will be further discussed in \S~\ref{sec:PeV}.

Of course, the possibility of analyzing spatially-resolved time-variations in the gamma-rays would provide essential clues to the acceleration mechanism, but this type of analysis is currently out of reach, due to the poor spatial resolution of the observations.  According to the picture discussed above, variations in the TeV domain are not expected to be dramatic in the case of Crab: being the emission mostly due to the interaction between radio emitting particles and internal synchrotron radiation \citep{Volpi:2008}, a radio-{\it wisp} like behaviour is expected, accompanied by very small variations of the integrated flux. However, the situation is completely different in the GeV range, where one is looking at the cut-off of the synchrotron spectrum and hence, in the case of Crab, at particles that have acceleration times comparable with radiation loss times. The dramatic consequences that this fact has on the Crab integrated emission in the GeV energy range will be the subject of next section.

\subsection{The Crab flares and their implications for particle acceleration}
\label{sec:flare}
A much unexpected discovery that came from gamma-ray observations of the Crab Nebula was that of episodes of extremely fast gamma-ray variability, the so-called gamma-ray flares. Global variations of the emissivity were predicted in the {\it Fermi} band as a consequence of rapid synchrotron burn-off of particles at the high energy cut-off of the distribution \citep{deJager:1996}. Assuming radiation reaction limited acceleration, the maximum energy up to which electrons can be accelerated is:
\begin{equation}
    E_{\rm max, rad}=m_e c^2 \left(\frac{6 \pi e \eta }{\sigma_T\ B}\right)^{1/2}\ \approx 6\ {\rm PeV}\ \eta^{1/2} B_{-4}^{-1/2},
    \label{eq:emaxrad}
\end{equation}
where $c$ is the speed of light, $e$ and $m_e$ are the electron charge and mass respectively, $\sigma_T$ is the Thomson cross section and we have assumed the acceleration to be due to an electric field $\eta\ B$, with $B$ the magnetic field strength. The second equality provides an estimate of the maximum achievable energy for magnetic field strengths in units of $B_{-4}=10^{-4}$ G, corresponding to the value estimated as the nebular average. One can easily see that PeV energies can only be reached for $\eta\approx 1$ and magnetic field strengths not much in excess of $10^{-4}$ G.

In this synchrotron-loss limited regime, it is easy to see that the maximum energy of synchrotron emitted photons only depends on $\eta$ and reads:
\begin{equation}
    \epsilon_{\rm max, sync}=\frac{3}{2}\hbar \frac{e B}{m_e c} \left(\frac{E_{\rm max,rad}}{m_e c^2}\right)^2=\eta \frac{9 \pi \hbar e^2 m_e c}{\sigma_T}\approx 230\ \eta\ {\rm MeV}\ .
    \label{eq:numaxsync}
\end{equation}

Global emissivity variations are therefore expected \citep{deJager:1996} in the hundreds of MeV range on time-scales:
\begin{equation}
t_{\rm var} \approx m_e\ c\ \sqrt{\frac{6 \pi}{e \sigma_T}}\ \eta^{-1/2} B^{-3/2}\approx 2.5\ \eta^{-1/2}\ B_{-4}^{-3/2}\ {\rm months}\ .
    \label{eq:tvarth}
\end{equation}

The big surprise came with {\it Agile} \citep{AGILEFlare} and {\it Fermi} \citep{FermiFlare} observations showing, on top of continuous small variations, some dramatic events, where not only the flux increases by a factor of several (up to 30 for the most spectacular event, in April 2011) over a period of 1-few weeks, but the emission extends well beyond $\epsilon_{\rm max,sync}$, reaching GeV photon energies. In addition, the amount of energy released is typically non-negligible, and in the biggest detected flare was really huge, corresponding to an isotropic luminosity of $L_{\rm max}= 4 \times 10^{36}$  erg/s $\approx 0.01 \dot E$. At present 17 flares have been clearly identified \citep{Huang21}, with a flare rate of 1.5 per year. In addition to episodes of sudden increase of the gamma-ray flux, also dips are observed in the same energy band \citep{BykovFlare20}.

The flares are not easy to interpret, and up to now there is still no accepted model to explain them. First of all, emission beyond 230 MeV implies $\eta>1$ which cannot be accommodated within ideal MHD. The possible solutions to this puzzle are: 1) the acceleration is due to a non-ideal mechanism with $\eta\gg1$, as can be the case for magnetic reconnection; 2) the acceleration occurs in a region of low magnetic field and then the emission occurs in a more magnetized region; 3) the emission comes from particles with mildly relativistic bulk motion, so that the frequency and power of the radiation are actually Lorentz boosted. All these possibilities have been widely explored in the literature.
In the first suggested scenario, acceleration of particles responsible for the flare would be part of the process of magnetic reconnection occurring in the vicinities of the TS. This idea has been thoroughly investigated by means of numerical simulations \citep{Cerutti:2012a,Cerutti:2013}. The general conclusion of these works is that acceleration by X-point magnetic reconnection would in fact explain emission beyond the synchrotron cut-off and a highly variable flux. In the brightest flare the flux doubles in less than 8 hr \citep{BlandBuehler}. Such a short time-scale implies emission from a very compact region, of size $L\approx 3\times 10^{-4}$ pc. In addition, if interpreted in terms of Eq.~\ref{eq:tvarth} implies $B\approx \eta^{-1/3}\ 3.7$ mG. Clearly this finding is challenging for any value of $\eta<1$, and in fact, as we will discuss later in more detail, it is challenging even for $\eta\approx 1$, in light of the recent LHAASO observations (see \S~\ref{sec:Others}). 

In a reconnection scenario the fast time-scale can be associated with the high level of fragmentation of the reconnection layer, made of a chain of magnetic islands, or plasmoids. Furthermore, these move with relativistic bulk speeds, which helps enhancing the intensity and frequency of the emitted radiation via Doppler boosting. Additional beaming is also provided by kinetic effects associated with the anisotropy of the particle distribution in the reconnection layer \citep{Cerutti:2014}. Despite all these promising features, 3D PIC simulations of magnetic reconnection indicate that the process is not fast enough to fully account for the properties of Crab flares \citep{Sironi_Cerutti:2017}: the reconnection rate is typically found to be found to be $v_{\rm rec}/c\lesssim 0.1$ \citep{Comisso2019}, likely translating into too weak an electric field.

A possible alternative is provided by explosive magnetic reconnection \citep{Nalewajko2016,Yuan2016,Lyutikov:2017a,Lyutikov:2017b}, where the process occurs on a dynamical time-scale. Very high Lorentz factors can be reached, because the highest energy particles are accelerated by the parallel electric field in the current layers and only suffer radiation losses after leaving the layer, building a scenario in which acceleration and radiation occur separately and the requirement $\eta>1$ imposed by Eq.~\ref{eq:emaxrad} is not an issue anymore. In addition the radiation is beamed, which helps with fast variability, and also with the implied energetics.

Besides scenarios invoking magnetic reconnection, a different class of models has attempted to explain the flares within the standard picture of Fermi acceleration. An early suggestion by \citep{Bykov12} is that the flare emission be interpreted as synchrotron emission in the cut-off regime in a magnetic field with stochastic fluctuations, such as is expected downstream of a shock that is efficiently accelerating particles. An interesting aspect of this picture is that it is proven to explain not only flux increases, but also depressions \citep{BykovFlare20}. The required magnetic field strength to explain the flare is in the mG range. The highly turbulent structure invoked by \citep{Bykov12,BykovFlare20} could be the outcome of another scenario that received much attention, that of a corrugated shock with mildly relativistic motion \citep{Lyutikov12,Lemoine16}. Of course the constraint from Eq.~\ref{eq:tvarth} would be relaxed if the variability has a different origin (unrelated to the acceleration time-scale) or if the emission comes from regions where the plasma is moving with a mildly relativistic speed, in which case, the intrinsic time-scale of the variations would be longer by a factor equal to the flow Lorentz factor. More recently, a modified picture of the shock, taking into account the latitudinal dependence of the magnetic field, has been numerically investigated \citep{CeruttiGiacinti20}, proving that mildly relativistic bulk motion develops, with Lorentz factor $\Gamma_w\sim 3-4$, enough to strongly relax all the constraints on frequency, time-variability and energetics of the flare. In particular, with $\Gamma_w$ in this range, also the previously discussed mechanism of ion-cyclotron absorption provides values in the right ballpark for the above mentioned quantities, even with a magnetic field around $100\,\mu$G.

\subsection{Constraints on the pulsar wind composition from >100 TeV emission}
\label{sec:PeV}
As discussed above, the pulsar wind is generally considered to be mostly composed of electron-positron pairs, while the possible presence of a hadronic component is still matter of debate \citep{Atoyan:1996,Bednarek:1997,Bednarek:2003,Amato:2003}.
If present, despite being a minority by number, hadrons could even be energetically dominant in the wind, changing drastically our understanding of the pulsar wind properties.
The relativistic hadrons possibly present in the Crab nebula could generate electromagnetic emission in the form of VHE gamma-rays deriving from decay of neutral pions produced in nuclear collisions with the gas in the SN ejecta. This spectral contribution is only expected to become detectable above $100\sim 150$ TeV, where IC scattering emission starts to be suppressed by the Klein-Nishina effect.

The current IACT (Imaging Atmospheric Cherenkov Telescopes), as H.~E.~S.~S. and MAGIC, could find no evidence of hadronic emission up to their sensitivity limit around tens of TeV.
Emission beyond 100 TeV is currently only accessible with sufficient sensitivity by water Cherenkov detectors and air shower detectors. Indeed, the Crab nebula was detected above 100 TeV by HAWC, employing the former technique \citep{HAWCcrab:2019} and by Tibet AS-$\gamma$ \citep{tibetASg:2019} employing the latter. 
Very recently LHAASO, combining both techniques, has obtained the record breaking detection of >PeV photons from this source \citep{cao2021peta}, opening up a window to finally see the possible emergence of the hadronic contribution. In fact, the increasing uncertainties above 500 TeV make the LHAASO spectrum still consistent with a purely leptonic origin of the emission. Under such an assumption, the PeV range data can be effectively used to constrain the strength of the magnetic field at the shock, which cannot exceed  $\bm{(112\, \pm 15)}\,\mu$G or otherwise, as one can readily see from Eq.~\ref{eq:emaxrad}, even assuming maximally efficient acceleration ($\eta$=1), radiation reaction would make it impossible to achieve particle acceleration up to the 2.8 PeV energy needed to explain the highest energy data point as due to IC scattering in the Klein-Nishina regime. A side remark is that in such a field, even a 2.8 PeV electron would emit synchrotron radiation at 50 MeV: even a Lorentz boost by a factor $\Gamma_w\sim 3-4$ would not be enough to account for the Crab gamma-ray flares. In other words the flares should come from a different region of the nebula, with higher magnetic field, or otherwise imply the presence of $\gtrsim 10$ PeV electrons, extremely close to the maximum potential drop available from the Crab pulsar, which is also the limiting energy for particles accelerated anywhere in the nebula (see \S~\ref{sec:Others}).

On the other hand, taken at face value, the LHAASO data seem to suggest that a new component might be showing up at the highest energies. This new component is consistent with a quasi-monochromatic distribution of protons with energy around 10 PeV (as discussed in Vercellone et al. in preparation). This is exactly what would be expected by models assuming that protons are part of the wind emanating from the Crab pulsar with a Lorentz factor $\Gamma_w\approx 10^7$: in this case their Larmor radius in a $100\,\mu$ G is of order $R_{\rm TS}$, so large that their energy distribution would not be much altered at the shock \citep{AmatoArons}. Of course smoking gun evidence would be the detection of neutrinos \citep{Amato:2003}, likely possible with the upcoming sensitivity improvement of dedicated experiments.

As far as gamma-ray data alone are concerned, in order to find clear evidence for the emergence of a hadronic component, more precise data and better modeling of the IC emission are needed, as well as a better understanding of the possible systematic entailed by the different techniques of VHE photon detection. The high altitude detectors provide flux measurements that are usually below those measured by IACT (compare H.~E.~S.~S. and MAGIC Crab data points with respect to Tibet As-$\gamma$ and LHAASO). While the discrepancy is not large, the error bars attached to the points do not overlap (see Fig.~\ref{fig:CRABspectrumIC}), which is somewhat puzzling, being the Crab nebula the primary calibration source in this energy range.
This lack of overlap might be due to systematic errors not being included in the error bars. 
On the other hand, multiple independent measurements of the Crab nebula spectrum in this energy range offer the perfect opportunity to properly asses the systematics of these complex observations. 
Decisive insight will be provided by next generation IACTs with good sensitivity beyond 100 TeV as the CTA SSTs (Small Size Telescopes) in the southern hemisphere and ASTRI Mini-Array in the north. 

Before concluding this section, we notice that the Crab nebula is not the only source to have been detected at EHE. 
Very recently LHAASO \cite{LHAASO_EHE} has also detected about ten more EHE emitters in the Galaxy (partially overlapping with the sources already detected by HAWC \cite{HAWC100} beyond 56 TeV). For the majority of these sources, the distance between the center of the emission and the nearest pulsar is less than or comparable with the instrument PSF, so that it is not unlikely that almost all these PeVatrons are associated with pulsars (and possibly leptonic in nature \citep{Breuhaus2021}). 
The much better spatial resolution of IACTs might also help to shed light on the real nature of these extreme accelerators, and assess whether acceleration of particles to PeV energies and beyond is a generic property of PWNe powered by energetic pulsars, rather than a unique property of Crab.

\section{The Crab nebula and the other PWNe}
\label{sec:Others}
While being considered as the prototype PWN, the Crab nebula is different from all other sources in this class in many respects, especially when it comes to gamma-rays. The first noticeable difference is that Crab is the only known PWN whose gamma-ray spectrum is partly formed with internal synchrotron radiation as a target. This is a consequence of its very bright synchrotron emission, due to the young age and high magnetic field. In addition, for the same reason, particle acceleration here is limited by radiation reaction, which is likely not the case for older objects with lower magnetic fields. In the latter, electrons can in principle be accelerated up to higher energies, comparable with the entire pulsar potential drop. In fact, the maximum achievable energy in the dissipation region, assumed to be located at a distance $R_{\rm TS}$ from the pulsar, is $E_{\rm max}= eB_{\rm TS} R_{\rm TS}$, where an electric field of the same strength as the magnetic field has been assumed. On the other hand, the magnetic field at $R_{\rm TS}$ can be estimated based on pressure equilibrium between the ram pressure dominated flow upstream of $R_{\rm TS}$ and the downstream: $B_{\rm TS}= \xi^{1/2} \sqrt{\dot E/c}/R_{\rm TS}$ with $\xi\leq 1$ the fraction of wind energy that is turned into magnetic energy. As a result: $E_{\rm max}\approx \xi^{1/2} e \sqrt{\dot E/c}$, namely a fraction $\xi^{1/2}$ of the pulsar potential drop, $E_{\rm drop}=e \sqrt{\dot E/c}$. 

The fact that in the majority of the observed PWNe, the maximum particle energy is not limited by radiation losses, might have something to do with the lack of flare observations from any source other than the Crab. It certainly has important implications for the escape of particles from evolved systems. At the same time, the fact that in evolved sources the VHE spectrum is uniquely due to upscattering of CMB photons (and occasionally local IR), has important consequences for the ratio between emission in different energy bands.
Particles responsible for the IC emission are generally less energetic than those responsible for the high energy synchrotron emission: a $\sim10$ TeV electron produces gamma-rays at 1 TeV, with the CMB as a target, while 1 keV synchrotron emission is produced by $\sim 50$ TeV electrons in an ambient magnetic field of 10 $\mu$G.
This difference in energy of the emitting electrons reflects in the different life time of a PWN in gamma-rays and X-rays, making the PWN to be still bright in gamma-rays when the X-ray emission is very low or even totally faded away.

Considering a rate of birth of 1 pulsar every 100 years in our Galaxy \citep{F-G-K06}, and an average lifetime of PWNe in gamma-rays of order of $100$ kyr, the total number of  PWNe possibly detectable at TeV energies is of order of $1000$. Most of these would be too old to be observed at other frequencies.
Evolved PWNe have in fact extended and diffuse radio emission, difficult to reveal on top of the background, while X-rays are hardly detectable due to the burn off of the emitting particle population.
Moreover old systems have gone through the so called \textit{reverberation} phase, when the SN reverse shock -- travelling towards the center of the SN explosion -- interacts with the PWN, likely causing a contraction of the nebula, with the consequent compression of the magnetic field and increase of the particle radiation losses  \citep{Gelfand:2009, Torres:2018, Bandiera:2020}.
Due to the system geometry and/or to the properties of the surrounding medium, the reverse shock is likely to be non-spherical and causes an asymmetric deformation of the nebula \citep{Temim:2015}. Additional deformation is likely induced by the PSR proper motion: the mean value of the kick velocity in the PSR population is of order $V_{\rm{PSR}}\sim 350$ km/s \citep{F-G-K06}, so that in a large fraction of sources, the pulsar will accumulated a sizeable displacement from the TeV emitting nebula during the system evolution.
The expected asymmetries and displacement from the parent pulsar position are then an important complication for the gamma-ray identification of PWNe.
As an example, out of the 24 extended sources revealed in the  H.~E.~S.~S. galactic plane survey \citep{HESS-GPS-2018}, only 14 have a multi-wavelength counterpart that allows for a firm association of the source with a PWN.
In the {\it Fermi}-LAT 3FGL catalog \citep{Acero:2015}, unidentified sources represent around $\sim 20\%$ of the detections at VHE.
It seems plausible that many of these unidentified bright gamma-ray sources are actually PWNe: the implication is  that this class can cover up to 40\% of the total gamma-ray sources in the sky.
A property of evolved PWNe that has attracted much attention in recent times is the release in the ISM of relativistic electron-positron pairs. This process has implications that go beyond PWN physics, since the pairs released by PWNs are currently the best candidates to provide an astrophysical explanation for the so-called {\it positron excess} observed in cosmic rays at energies above $\sim$10 GeV \citep{Adriani:2009,Aguilar:2013,Bykov:2017}.

The most energetic particles in the nebula, with energy close to $E_{\rm drop}$, have been shown to efficiently escape from the head of the bow shock that forms at the interface between the PWN and the ISM, once the pulsar has emerged from the SNR ({\it bow-shock} PWN).
Those particles have large Larmor radii, comparable with the bow shock thickness in the head of the system, and can stream in the outer medium along the magnetopause at the contact discontinuity between the nebula and the ISM \citep{Olmi&Bucciantini:2019_3}.
\begin{figure*}
\centering
	\includegraphics[width=.8\textwidth]{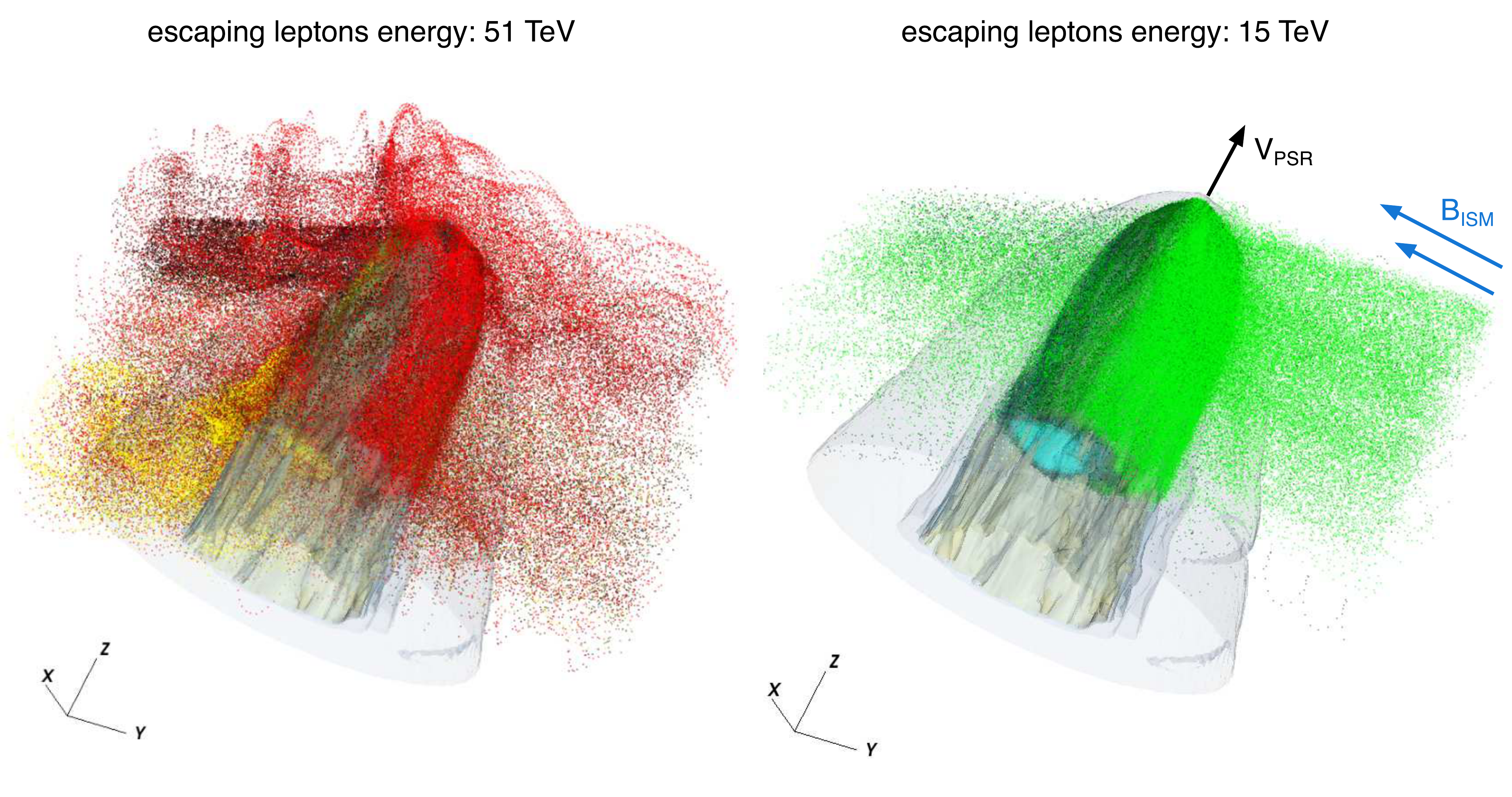} 
    \caption{Maps of bow shock nebulae from 3D MHD simulations with density contours (in grey color) and the flux of escaping leptons (of two different energies). Dots of different colors indicate particles injected at different locations in the pulsar wind: the majority of escaping particles are injected in the polar region of the wind (red and green), while very few of them come from the equatorial region. In both plots the PSR direction of motion is aligned with the $Z$ direction, while the magnetic field, with strength $B_{\rm{ISM}}=0.01 \rho_{\rm{ISM}}V_{\rm{PSR}}^2$ (with $\rho_{\rm{ISM}}$ the ISM mass density) lies in the orthogonal plane. Plots have been elaborated based on the simulations presented in \citep{Olmi&Bucciantini:2019_3}.\\
    The figure on the left is reprinted from Olmi \& Bucciantini 2019 $\copyright$ 2019 Olmi \& Bucciantini.}
    \label{fig:bsesc}     
\end{figure*}
Depending on their energy and on the properties of the surrounding ISM, the escaping particles are expected to form diffuse halos around the bow shock head or extended and collimated jets, eventually misaligned with the pulsar direction of motion (see Fig.~\ref{fig:bsesc}) and somehow similar structures have been observed in the last years to emerge from many bow shock nebulae in the X-rays \citep{Hui:2007,Pavan:2014,Temim:2015,deVries:2020,Kim:2020,Zhang:2020,Wang:2021}.
The escaping particle flux also shows evidence of effective charge separation.

This property could play a key role to understand the formation of the so-called {\it gamma-ray halos}. This new class of sources was first identified by HAWC, which detected extended halos of multi-TeV emission surrounding two evolved systems: Geminga (PSR B0633+17) and the Monogem (PSR B06556+14)  \citep{Abeysekara:2017}. 
The size of the halo around Geminga is much larger than the observed size of the nebula in X-rays ($\sim 25$ pc vs $\sim0.2$ pc), so that it must be produced by particles that have escaped from the system. On the other hand, the extension is too small to be produced by particles that propagate in the standard Galactic diffusion coefficient, since the expected size would be a factor of $\sim100$ larger in that case.
A possible explanation has been searched for in a modification of the diffusion properties around that source, possibly conveyed by self-generated turbulence associated with electrons and positrons leaking the nebula \citep{Evoli:2018}, or due to the injection of MHD turbulence by the parent SNR \citep{Fang2019}. At present, understanding the formation of TeV halos is one of the big challenges in high energy astrophysics (see e.g. Lopez-Coto et al. in preparation), both for their possible implications for galactic cosmic ray transport and for their implications for future gamma-ray observations. In fact, these could provide an important source of confusion, being weak and extended and not easy to identify.
The number of expected detectable halos in the TeV sky is also still matter of  debate, with estimates ranging from very many ($\sim50-240$ \citep{Sudoh:2019} to a few \citep{Giacinti:2020}.
The need for better theoretical understanding and physically motivated predictions of their abundance and location is apparent.

In this respect, the Crab nebula is certainly not a prototype, and a better understanding of this source is doubtful to help.

\section{Summary and future prospects}
\label{sec:sumEprosp}
The Crab nebula and its pulsar are certainly among the most studied astrophysical sources in the sky, and as such they provide an excellent laboratory to investigate many aspects of high energy astrophysics and relativistic plasma physics. At the same time, this system has proven to be an endless source of surprises.
The discovery of the Crab pulsar was the confirmation that radio pulsars are actually rotating neutron stars, while the study of the Crab nebula has taught us that most of the pulsar spin-down energy goes into a highly relativistic and magnetized outflow. In this article we reviewed what we have learned about the pulsar and the nebula in the last two decades. While both objects have a very broad emission spectrum, high energy observations, and gamma-ray observations in particular, have played a special role in recent developments. 

We have seen in \S~\ref{sec:PSR} how HE and VHE observations have put stringent constraints on the origin of pulsed gamma-ray emission, enforcing the view that it is produced far from the pulsar, at distances $\gtrsim R_{LC}$, and suggesting new scenarios for the related process of pair-creation. 

In \S~\ref{sec:PWN} we have reviewed how our understanding of the PWN plasma dynamics has changed in recent years, thanks to a combination of improved modeling and high quality observations. We have discussed how 2D and 3D MHD models of the nebular dynamics have allowed to solve (or alleviate) some of the mysteries of the Crab nebula, like the {\it wisps} activity, the origin of the X-ray emitting jet or the $\sigma$-problem. The jet is explained as a result of an anisotropic energy flow from the pulsar (higher along the pulsar rotational equator than along the polar axis) and the dynamical effect of the hoop stresses associated with the toroidal magnetic field. This explanation requires the wind magnetization $\sigma$ to be sufficiently large. The latter must be much larger than the value of $\sigma\sim 10^{-3}$ that was originally estimated, and likely $\sigma\gtrsim$ a few, in order for the gamma-ray spectrum of the nebula to be correctly accounted for. The variability of the {\it wisps} is naturally found in time-dependent MHD modeling, and the {\it wisps} appearance at different wavelengths implies different locations for the acceleration of particles in different energy ranges: in particular X-ray emitting particles must be accelerated in the equatorial sector of the shock, while lower energy particles can be accelerated anywhere. What mechanisms are responsible for particle acceleration in the different energy ranges is an unsettled question, because all the proposed mechanisms have strengths and weaknesses, and none can be completely ruled out for lack of better knowledge of the wind composition and magnetization at different locations along the shock front.  

 In spite of our ignorance of what process is actually at work, in \S~\ref{sec:flare} and \ref{sec:PeV} we have showed how extraordinary an accelerator the Crab nebula is, as highlighted in the last decade by the gamma-ray flares and, very recently, by the detection of PeV photons. Several different scenarios have been proposed to explain the flare, with its emission beyond the synchrotron cut-off frequency and extremely fast variability. 
 However most of these proposals assume the emission to come from a region with mG strength magnetic field. Such a value of the field is one order of magnitude larger than implied by the detection of PeV emission, if this is of leptonic origin and due to IC scattering. 
 
 The PeV data are also especially intriguing because there is a suggestion that a new component might be showing up at the VHEs, consistent with a quasi-monochromatic distribution of protons with energy $\sim 10$ PeV. The presence of hadrons in the pulsar wind would be a paradigm changing discovery: not only it would change the current view of the pulsar outflow (with effects on the modeling of both the pulsar magnetosphere and the nebula) but it would also have consequences on cosmic ray astrophysics, lending support to the idea that fast spinning, highly magnetized neutron stars can be major contributors of ultra high energy cosmic rays.
 
 However, as we discussed in \S~\ref{sec:PeV}, smoking gun evidence for the presence of hadrons in the Crab pulsar wind requires more precise data and possibly better control on the systematics at VHE. The contribution of hadrons in the Crab spectrum is expected to emerge above around $150-200$ TeV, where IC starts to be suppressed by the Klein-Nishina effect. The next generation of IACTs (CTA and ASTRI Mini-Array), with sensitivity extended to this energy range, is likely to play a crucial role in finally answering this question.

As we discussed in \S~\ref{sec:Others}, the gamma-ray astronomy community has long been much interested in PWNe as the dominant class of galactic sources, and this interest has been recently increased by the discovery of gamma-ray halos around pulsars. The advent of the new generation of high sensitivity and high resolution IACTs, with special reference to CTA, will give us access to a huge amount of new data. PWNe will be the largest population of gamma-ray sources in future surveys (possibly up to 40\% of the total). The expected number of newly detected PWNe by CTA is of order 200, while the number of detectable halos is right now very uncertain.

In terms of the population of gamma-ray emitting PWNe, the Crab cannot be considered as prototypical: due to its young age and high magnetic field the Crab is, in fact, the known PWN whose IC spectrum is partly due to self-synchrotron radiation and one of the few gamma-ray emitting PWNe in which the maximum particle energy is determined by radiation losses, rather than shortage of available potential. Especially the latter condition is critical in determining the presence or absence of a halo, since only particles close to the maximum pulsar potential drop are expected to efficiently escape from the nebula and form a IC scattering halo. Based on available simulations, efficient particle escape at lower energy is only possible from the tail of pulsar Bow shock nebulae. This is an important aspect to assess quantitatively in view of explaining the cosmic ray positron excess as due to pulsars. Measurements of the total lepton spectrum at VHE, which will be possible with next generation IACTs, will contribute to clarify this issue. On the other hand, more refined modeling of the highest energy particle escape and associated plasma instabilities should help clarify the nature of gamma-ray halos and their expected abundance. These are again crucial problems also for cosmic ray physics, since they could imply a change of our description of particle transport in the Galaxy. At the same time, the detectability of gamma-ray halos, as well as that of evolved PWNe is a major challenge for gamma-ray astronomy, since these weak and extended sources not only are scientifically interesting, but also need to be taken into account carefully as back-ground contributors against the detection of other sources, most notably potential hadronic PeVatrons, whose identification is one of the main science goals of upcoming facilities. 

Going back to Crab, this source is very different, in many respects from the evolved PWNe that future IACTs will detect in very large numbers. In this sense, the Crab is not the source to look at if the purpose is that of learning about the average properties of gamma-ray emitting PWNe. On the other hand, the Crab keeps being the best place to learn about the processes that make these objects such extreme accelerators, both in terms of efficiency and achievable energies. By looking at this ever surprising source, future IACTs might be able to tell us that PWNe are themselves hadronic PeVatrons.

\funding{This research was funded by  Italian Space Agency (ASI) and National Institute for Astrophysics (INAF) under the agreements ASI-INAF n.2017-14-H.0, from INAF under grants:  “PRIN SKA-CTA”, “INAF Mainstream 2018” and “PRIN-INAF 2019”.}

\acknowledgments{We acknowledge Rino Bandiera and Niccolò Bucciantini for continuous collaboration during the years, and for providing useful comments on this manuscript. We also thank Michele Fiori for providing us with the updated gamma-ray spectrum of the Crab Nebula shown in Figure \ref{fig:CRABspectrumIC}.}

\abbreviations{The following abbreviations are used in this manuscript:\\
\noindent 
\begin{tabular}{@{}ll}
ASTRI & Astrofisica con Specchi a Tecnologia Replicante Italiana \\
CMB & Cosmic Microwave Background \\
CTA & Cherenkov Telescope Array \\
{\it Fermi}-LAT &  {\it Fermi}-Large Area Telescope\\
EHE & Extremely High Energy  \\
HD & HydroDynamic \\
HE & High Energy  \\
H.E.S.S. & High Energy Stereoscopic System \\
IACT & Imaging Atmospheric Cherenkov Telescope \\
IC & Inverse Compton  \\
ISM & InterStellar Medium \\
LHAASO & Large High Altitude Air Shower Observatory\\
MAGIC & Major Atmospheric Gamma-ray Imaging Cherenkov\\
MHD & Magneto-HydroDynamic \\
PIC & Particle In Cell \\
PSR & Pulsar \\
PW & Pulsar Wind  \\
PWN & Pulsar Wind Nebula  \\
SN & Supernova \\
SNR & Supernova Remnant  \\
VERITAS & Very Energetic Radiation Imaging Telescope Array System  \\
VHE & Very High Energy  \\
\end{tabular}
}

\reftitle{References}

\externalbibliography{yes}
\bibliography{biblio.bib}

\end{document}